\documentclass[twocolumn,showpacs,preprintnumbers,amsmath,amssymb]{revtex4}
\usepackage[dvipdfmx]{graphicx}
\usepackage{dcolumn}
\usepackage{bm}
\usepackage{amssymb}
\usepackage{amsmath}
\usepackage{color}
\usepackage{natbib,hyperref}

\begin{document}

\title{Bilayer sheet protrusions and budding from bilayer membranes induced by hydrolysis and condensation reactions}% Force line breaks with \\

\author{Koh M. Nakagawa and Hiroshi Noguchi}
 \email{noguchi.issp.u-tokyo.ac.jp}
\affiliation{Institute for Solid State Physics, University of Tokyo, Kashiwa, Chiba 277-8581, Japan}%

\date{\today}% It is always \today, today,
             %  but any date may be explicitly specified

\begin{abstract}
  Shape transformations of flat bilayer membranes and vesicles induced by hydrolysis and condensation
reactions of amphiphilic molecules are studied using coarse-grained
molecular dynamics simulations. The hydrolysis and condensation reactions result in
the formation and dissociation of amphiphilic molecules, respectively.
Asymmetric reactions between the inner and outer leaflets of a vesicle can transport
amphiphilic molecules between the leaflets.
It is found that the resulting area difference between the two leaflets induces
 bilayer sheet protrusion~(BP) and budding at low reduced volumes of the vesicles,
whereas BP only occurs at high reduced volumes.
The probabilities of these two types of transformations
depend on the shear viscosity of the surrounding fluids
compared to the membrane as well as the reaction rates.
A higher surrounding fluid viscosity leads to more BP formation.
The inhomogeneous spatial distribution of the hydrophobic reaction products
forms the nuclei of BP formation, and faster diffusion of the products enhances BP formation.
Our results suggest that adjustment of the viscosity is important
to control membrane shape transformations in experiments.
\end{abstract}

\pacs{87.10.Tf,83.10.Rs,87.16.D-}% PACS, the Physics and Astronomy
                             % Classification Scheme.
%\keywords{}
\maketitle

\section{Introduction}
A lipid vesicle, which is one of the basic self-assembled structures of lipid
molecules, has been studied as a minimum model of living cells. 
Although vesicles are composed of only lipid molecules, they exhibit various shape
transformations such as discocyte, stomatocyte, and starfish depending on
reduced volumes and spontaneous curvatures~\cite{seifert1997configurations}.
These various shape transformations can be well described using the elastic
theory proposed by Helfrich~\cite{helfrich1973elastic,seifert1997configurations}.
The theoretical prediction agrees well with experiments quantitatively~\cite{sakashita2012three}.

For the above-mentioned studies on vesicle morphology,
it is assumed that the membrane composition is constant. 
However,
in living cells, synthesis and decomposition of lipids
 continually occur by lipid metabolism
so that the membrane composition changes. 
For example, phospholipids are synthesized from fatty
acids on endoplasmic reticulum (ER) membrane~\cite{bell1981lipid}, after which these molecules
are transported to other organelles via vesicle transport mechanisms.
Another example is hydrolysis of phospholipids~\cite{nishizuka1992intracellular,nishizuka1984role}.
The reaction products, diacylglycerols (DAGs), play a key role in
protein kinase C activation~\cite{bell1986protein}.

The effects of non-constant membrane composition on shape transformations of
cells have been studied by several
groups~\cite{allan1975changes,allan1978rapid,ibarguren2013recruitment,holopainen2000vectorial,riske2003diacylglycerol}.
The nature of non-constant membrane
composition is often mimicked by hydrolysis and condensation reactions. 
For example, an injection of a hydrolase enzyme to
red blood cells~(RBCs)~\cite{allan1975changes,allan1978rapid} and lipid
vesicles~\cite{ibarguren2013recruitment,holopainen2000vectorial} hydrolyzes
amphiphilic molecules into hydrophilic and hydrophobic molecules.
In these chemical reactions, amphiphilic molecules, which are
composed of hydrophilic (A) and hydrophobic (B) parts, change as follows:
\begin{eqnarray}
  \mathrm{AB} + \mathrm{H}_{2}\mathrm{O} \rightleftharpoons \mathrm{A}\mathrm{-H} + \mathrm{B}\mathrm{-OH}, \label{eq:chem}
\end{eqnarray}
where the right and left arrows represent the hydrolysis and condensation reactions, respectively.
The experiments demonstrate that vesicles show the various shape transformations
under these chemical reactions~\cite{allan1975changes,allan1978rapid,ibarguren2013recruitment,holopainen2000vectorial}. 
The injection of hydrolase enzyme to
RBCs and liposomes induces membrane invagination and rupture.
Toyota {\it et al.}~\cite{toyota2006hierarchical} reported that discocytes
transform into tubular and invaginated shapes under the hydrolysis reaction
of amphiphilic molecules.

A possible explanation of these morphological changes was proposed in terms of 
the area-difference-elasticity~(ADE) model~\cite{allan1978rapid}.
The hydrolase enzyme is injected from the outer solution
so that the density of the amphiphilic molecules in the outer leaflet of a vesicle
decreases, whereas the density of the inner leaflet is nearly
constant. Therefore, the amphiphilic molecular densities in the inner and outer leaflets
become different. To reduce the ADE energy~\cite{sheetz1974biological,seifert1997configurations} 
of this area difference,
liposomes and RBCs form an invagination.
A similar asymmetry in amphiphilic molecules that is induced by chemical
reactions is widely observed {\it in vitro}~\cite{suzuki2009sparkling} and
{\it in vivo}~\cite{bell1981lipid}.

However, in the previous works,
the effects of the resulting products (A--H and B--OH molecules in eqn~(\ref{eq:chem})) 
are not taken into account explicitly. Instead, the effect of these chemical
reactions is taken into account implicitly by the change in the amphiphilic
molecular densities of the inner and outer leaflets in the ADE model.
The resulting hydrophobic molecules (B--OH) are included in the bilayer
membrane, and such inclusions modify the elastic properties of the
bilayer~\cite{ly2004influence}. We have previously examined how such inclusions
affect shape transformations from an oil droplet to a bilayer membrane
by a binding reaction of hydrophobic and hydrophilic molecules~\cite{nakagawa2015morphological}.
Tubular vesicles form via bicelles. 
The inclusions are concentrated in the branches of the membranes.
The stabilization of branched structures by inclusions was also reported in
Refs.~\cite{das1986modification,basanez1997morphological}.
Thus, shape transformations of membranes occur due
to the presence of the resulting hydrophobic molecules, but these shape
transformations are not fully understood.

The aim of this paper is to clarify the effect of embedded hydrophobic
products on shape transformations under the hydrolysis and condensation
reactions. We use the coarse-grained molecular dynamics simulation technique,
in which the hydrolysis and condensation reaction processes of amphiphilic
molecules are taken into account explicitly. We will show that the shape transformation strongly depends on the
distribution of the resulting products (B--OH) included in the bilayer.
We will also show that not only membrane invagination but also bilayer sheet
formation occur depending on the reduced volume.
The shape transformation pathway is also affected by the
transport coefficients of the surrounding fluids and membrane.

In Section~\ref{sec:mod},
 the simulation model, method, and simulation settings are described. 
The results are presented in Section~\ref{sec:mor}, and discussions and
conclusions are given in Section~\ref{sec:conc}.

\section{Simulation methods}\label{sec:mod}
\subsection{Model and method}
We use the dissipative particle dynamics (DPD) simulation
technique~\cite{groot1997dissipative, hoogerbrugge1992simulating, groot2001mesoscopic}.
In the DPD method, the particle motion is given by the following
Newton's equations with a pairwise Langevin thermostat:
 \begin{align}
  \label{eq:dpd}
  m \dfrac{d{\bf v}_{i}}{dt} &= -\dfrac{\partial U}{\partial {\bf r}_{i}} \\
  &+ \sum_{j \neq i} \left(-w(r_{ij}) {\bf v}_{ij} \cdot \hat{{\bf r}}_{ij} + \sqrt{w(r_{ij})}\xi_{ij}(t)\right) \hat{{\bf r}}_{ij}, \nonumber 
 \end{align}
with
\begin{align}
 &U = \sum_{i>j} U_{\mathrm{rep}}(r_{ij}) + \sum_{\mathrm{bonds}} U_{\mathrm{bond}}(r_{ij}) + \sum_{\mathrm{angles}} U_{\mathrm{angle}}(\theta_{ijk}), \\ \label{eq:wgt}
 &w(r_{ij}) = \gamma_{ij} \left(1 - \dfrac{r_{ij}}{r_{\mathrm{cut}}} \right)^{p}\Theta \left(1 - \dfrac{r_{ij}}{r_{\mathrm{cut}}} \right),
\end{align}
where ${\bf v}_{ij} = {\bf v}_{i} - {\bf v}_{j}, {\bf r}_{ij} = {\bf r}_{i} - {\bf r}_{j}, r_{ij} = |{\bf r}_{ij}|, \hat{{\bf r}}_{ij} = {\bf r}_{ij} / r_{ij}$,
and $\Theta$ is the unit step function. A harmonic potential is used for the repulsive potential,
{\it i.e.}, $U_{\mathrm{rep}}(r_{ij}) = a_{ij}(1 - r_{ij}/r_{\rm {cut}})^{2}/2$,
which vanishes at the finite cutoff $r_{\mathrm{cut}}$.
The Gaussian white noise $\xi_{ij}(t)$ satisfies the fluctuation and dissipation theorem, {\it i.e.},
$\langle \xi_{ij}(t) \rangle = 0, \langle \xi_{ij}(t) \xi_{kl}(t') \rangle =
2k_{\mathrm B}T(\delta_{ik}\delta_{jl} + \delta_{il}\delta_{jk}) \delta (t-t')$,
where $k_{\mathrm B}T$ is the thermal energy.
An amphiphilic molecule consists of hydrophilic head and hydrophobic tail segments that
 are represented by one and three particles, respectively.
These four particles are connected via the bond potential $U_{\mathrm{bond}}(r_{ij}) = k_{\mathrm{bond}}(1 - r_{ij}/l)^{2}/2$ and the angle potential
$U_{\mathrm{angle}}(\theta_{ijk})  = k_{\mathrm{angle}}(1 - \cos \theta_{ijk})$ with
$k_{\mathrm{bond}}=272k_{\mathrm B}T$, $k_{\mathrm{angle}}=60k_{\mathrm B}T$, and $l=0.8r_{\mathrm{cut}}$.

Many DPD simulations~\cite{groot1997dissipative, venturoli2006mesoscopic} are carried out with $p=2$ in eqn~(\ref{eq:wgt})
 to reduce computational costs, but a liquid phase
(Schmidt number $Sc \gtrsim 10$) is not obtained without increasing
$\gamma_{ij}$ up to
50$\sqrt{m k_{\mathrm B}T}/r_{\mathrm{cut}}$~\cite{noguchi2007transport}.
Instead, we choose $p=1/2$ to increase the shear viscosity of the DPD fluids~\cite{fan2006simulating}.
We use the Shaldrow S1 splitting algorithm~\cite{shardlow2003splitting} to
discretize eqn~(\ref{eq:dpd}).
The multi-time-step algorithm~\cite{peters2004elimination, tuckerman1992reversible, noguchi2007transport}
is employed with the integration time step $\Delta t = 0.005 r_{\mathrm{cut}} \sqrt{m / k_{\mathrm B}T}$
for the conservative forces and $\delta t = 0.05 r_{\mathrm{cut}} \sqrt{m / k_{\mathrm B}T}$ for
dissipation and random forces.

The repulsive parameters $a_{ij}$ are listed in Table \ref{tb:repul}. The dissipative parameters
$\gamma_{ij}$ for the same type of particle pairs are shown in Section~\ref{sec:mor}.
For different types of particle pairs, a harmonic mean rule is employed, {\it i.e.},
$\gamma_{ij} = 2 / (1/\gamma_{ii} + 1/\gamma_{jj})$, to ensure
the correct hydrodynamic behavior for the multi-viscosity system~\cite{visser2006modelling}.
A stable bilayer structure forms with bending 
rigidity $\kappa=18.1\pm0.4
k_{\mathrm B}T$ and area expansion
modulus $K_{A}=18.9\pm1.1k_{\mathrm B}T/r_{\mathrm{cut}}^{2}$.
These elastic properties agree well with the experimental
results at room temperature~\cite{seifert1997configurations,boal2012mechanics}.

\begin{table}
 \begin{center}
  \begin{tabular}{crrr}\hline
      & W  & H  & T   \\ \hline
     W  & 25 & 25 & 200 \\
     H  & 25 & 25 & 200 \\
     T  & 200& 200& 25 
  \end{tabular}
   \caption{Repulsive interaction parameters $a_{ij}$ with units
     $k_{\mathrm{B}}T$. W, H, and T represent water, hydrophilic and hydrophobic
     particles, respectively. \label{tb:repul}}
 \end{center}
\end{table}

We use reduced units with $r_{\mathrm{cut}}$ as the unit of length, $k_{\mathrm B}T$ 
as the unit of energy, and $m$ as the unit of mass.
$r_{\mathrm{cut}}$ is the length scale of the molecule, $r_{\mathrm{cut}} \sim 1$ nm, and $\tau = r_{\mathrm{cut}}\sqrt{m/k_{\mathrm{B}}T}$ is estimated to be
$\sim$1 ns at room temperature $T \sim 300$ K.
Dimensionless quantities are denoted by $^{*}$, {\it e.g.}, $t^{*}=t/\tau$.

\begin{figure}[htbp]
   \centering
   \includegraphics[width=60mm]{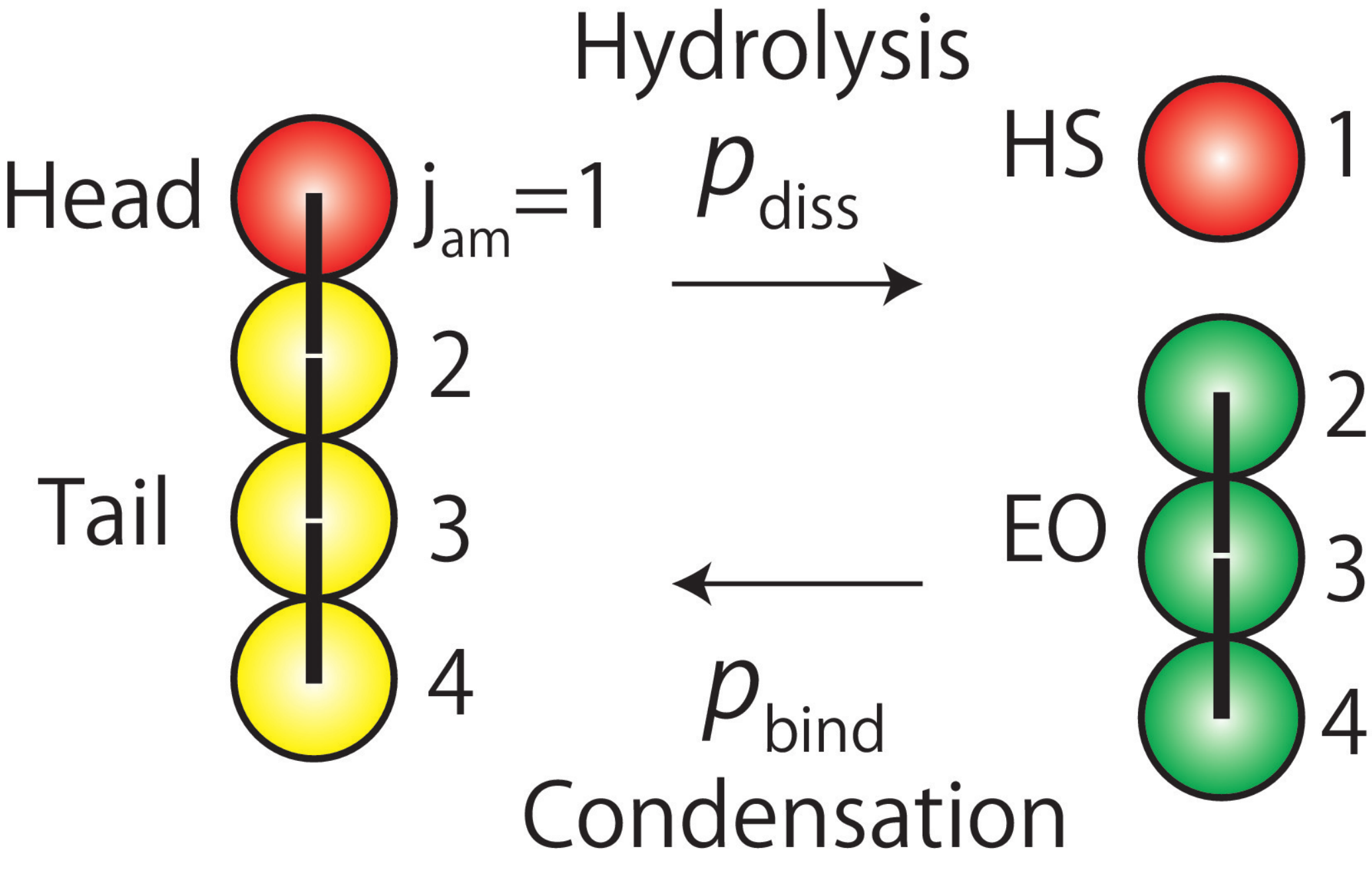}
   \caption{Schematic picture of hydrolysis and condensation reactions. 
     For clarity, the hydrophobic particles in amphiphilic molecules and
     in the EOs are colored yellow and green, respectively.
     Both hydrophilic and hydrophobic particles have their own ids: 
     $j_{\mathrm{am}}=1$ for the hydrophilic particles,
     and $j_{\mathrm{am}}=2, 3, 4$ for the hydrophobic particles.}
   \label{fig:chem_schem}
  \end{figure}

Several chemical reaction models combined with the DPD method have been
proposed~\cite{lisal2006mesoscale,lisal2009mesoscale,huang2016insight,nakagawa2015morphological}.
In this work, we use the chemical reaction model, in which
hydrolysis and condensation reactions of amphiphilic molecules are represented by
a bond dissociation and bond binding as shown in Fig.~\ref{fig:chem_schem}.
Because the dissociated hydrophilic and hydrophobic molecules are typically
dissolved in surrounding fluids and embedded in the bilayer,
we refer to them as the hydrophilic solute (HS) and embedded oil (EO), respectively.
The HS can have a binding reaction to only the one end particle of the EO ($j_{\mathrm{am}}=2$ shown in Fig.~\ref{fig:chem_schem}).
Both binding and dissociation processes are
treated as stochastic processes, as in the polymerization
model~\cite{huang2016insight}.
Probabilities for the bond binding and dissociation
during $\Delta t$ are given by
 \begin{align}
  p_{\mathrm{diss}}  &=
  \begin{cases}
   p_{\mathrm{f}} \Delta t & (n_{\mathrm{water}} > 0)\\
   0 & (\mathrm{otherwise})
  \end{cases}, \\ 
  p_{\mathrm{bind}} &=
  p_{\mathrm{r}} \Delta t \Theta \left(1 - \dfrac{r_{\mathrm{min}}}{r_{\mathrm{bind}}} \right),
 \end{align}
where $p_{\mathrm{f}}$ and $p_{\mathrm{r}}$ denote
the transition rates
of the dissociation and binding reactions, respectively.
 $n_{\mathrm{water}}$ is the
number of water particles that exist in a sphere with a radius of $0.69r_{\mathrm{cut}}$
around a hydrophobic particle of $j_{\mathrm{am}}=2$ connecting with a hydrophilic particle.
The hydrophobic particles of $j_{\mathrm{am}}=2$ in the EO
bind with the closest HS by the reaction rate $p_{\mathrm{r}}$
when the distances $r_{\mathrm{min}}$ between the two particles are less than
the cutoff length $r_{\mathrm{bind}}$. In this study, $r_{\mathrm{bind}}=r_{\mathrm{cut}}$ is used.
The bond dissociation probability relatively increases when the HS concentration is low.
When the bond binding rate is equal to the bond dissociation rate,
the system reaches chemical equilibrium.

We investigate the effects of the viscosity $\eta_{\mathrm{sol}}$ of the surrounding solutions
and the effective viscosity  $\eta_{\mathrm{mb}}$ in the membrane.
The viscosity of the  surrounding fluids can be controlled 
by $\gamma_{\mathrm{sol}} = \gamma_{\mathrm{HH}} = \gamma_{\mathrm{WW}}$.
The shear viscosity of the DPD fluids increases with increasing $\gamma$.
The shear viscosity $\eta_{\mathrm{sol}}$
is estimated from a simple shear flow for the DPD fluid 
consisting of only W particles~\cite{noguchi2007transport}:
$\eta_{\mathrm{sol}}^{*}=1.119 \pm 0.004$ and $5.262$ $\pm$ 0.007 for $\gamma_{\mathrm{sol}}^{*} = 2$ 
and $24.5$, respectively.
The diffusion constants $D$ of the DPD fluids are estimated  from the mean square
displacement: $D^{*}=1.326$ and $0.301$ for $\gamma_{\mathrm{sol}}^{*} = 2$ and $24.5$, respectively.
To change the effective viscosity $\eta_{\mathrm{mb}}$ in the membrane,
we vary $\gamma_{\mathrm{TT}}$ of the tail and EO particles: $\gamma_{\mathrm{TT}}^{*} = 2$ and $\gamma_{\mathrm{TT}}^{*} = 24.5$.
Thus, the DPD fluids consisting of nonbonded tail particles have $\eta_{\mathrm{mb}}^{*}=1.119$ and $5.262$.
However, the bond and angle potentials in the amphiphilic molecules modify this simple linear viscosity.
In the bilayer membranes, 
the amphiphilic molecules in the membrane have two types of hydrodynamic interactions: 
lateral interactions that give rise to 2D membrane viscosity and the friction between
two leaflets~\cite{seifert1997configurations,den2007intermonolayer,shkulipa2005surface}.
A greater value of $\gamma_{\mathrm{TT}}$ yields a higher 2D viscosity and stronger friction.
However, because the EOs in the bilayer modify both the interactions,
it is difficult to quantitatively estimate them.
Therefore, we consider only the qualitative effects of the membrane viscosity 
using the viscosity $\eta_{\mathrm{mb}}$ of the DPD fluid, in this study.

\subsection{Simulation settings}\label{sec:simset}
All simulations are carried out in an $NVT$ ensemble (constant number $N$ of
particles, volume $V$, and temperature $T$)
at the particle density $N/V = 3/r^{3}_{\mathrm{cut}}$.
The cubic simulation boxes with $L_{x} = L_{y} = L_{z} = 36r_{\mathrm{cut}}$ and $48r_{\mathrm{cut}}$ are
used for a flat membrane and vesicle, respectively.

We prepare a flat bilayer membrane with  $N_{\mathrm{amp}} = 4950$ and $N_{\mathrm{EO}} = 0$.
The bounce-back rule is employed at the boundary of the simulation box along the normal ($z$) direction to the bilayer, and
periodic boundary conditions are employed in the lateral ($x,y$) directions.
Initially, the HS concentrations of the fluids above and below the membrane are
 $c^{*} = 3$ and $0$, respectively. Since these two fluids do not contact each other directly,
their HS concentrations are changed only by the reactions on the membrane.

To investigate the spatial distribution of the EOs in the membrane in the absence of the chemical reactions,
a flat membrane with $N_{\mathrm{EO}}=1000$ and $N_{\mathrm{amp}} = 3950$ is used.
The EOs are distributed uniformly in the bilayer membrane as initial
conformations, and the positions of the EOs are fixed
for the first $500 \tau$ to make the amphiphilic molecules relax first.
The position constraint of the EOs is removed at $t = 500 \tau$.
The surface density is calculated with a spatial mesh size of $r_{\mathrm{mesh}} =
4r_{\mathrm{cut}}$.

We consider initially a nearly spherical vesicle and a discocyte-shaped vesicle
under periodic boundary conditions in all three directions.
The spherical vesicle is formed by $N_{\mathrm{amp}} =17730$ 
of amphiphilic molecules, and  $N_{\mathrm{in}} = 52670$ particles
are inside the vesicle. The discocyte has $N_{\mathrm{amp}} = 19338$ and $N_{\mathrm{in}} = 35679$.
The reduced volumes $v=V_{\mathrm{ves}}/(4/3\pi(A_{\mathrm{mb}}/4\pi)^{3/2})$
for the spherical vesicle and discocyte
are nearly equal to 1 and 0.5, respectively,
where  $V_{\mathrm{ves}}$ and $A_{\mathrm{mb}}$ are the volume
and surface area of the membrane.
Initial vesicles are prepared using the methods described in Appendix \ref{apdx:prep_ves}.
The simulation time $t$ is set to zero when the chemical reactions start.

In the experimental
studies~\cite{allan1975changes,allan1978rapid,ibarguren2013recruitment,holopainen2000vectorial}
shown in Section 1, the chemical
environment is different inside and outside a vesicle.
As a model of such asymmetric situations, 
we consider the vesicle whose inner and outer solutions have different concentrations of HSs.
One may consider that the asymmetric concentration along the bilayer membrane
causes the osmotic pressure difference.
However, the time scale of the volume change due to the osmotic pressure difference
is much longer than the shape transformation timescale.
We will show how the shape
transformation of vesicles under the hydrolysis and condensation reactions
is changed by the concentration difference of HSs.
The initial concentration $c_{\mathrm{out}}$ of HSs outside the vesicle
is set to 0. The initial concentration $c_{\mathrm{in}}$ of HSs inside the vesicle is varied to
control the concentration difference.

\section{Simulation results}\label{sec:mor}
\subsection{Bilayer sheet protrusion from flat membrane}\label{sec:flat}
\begin{figure}[htbp]
 \centering
 \includegraphics[width=70mm]{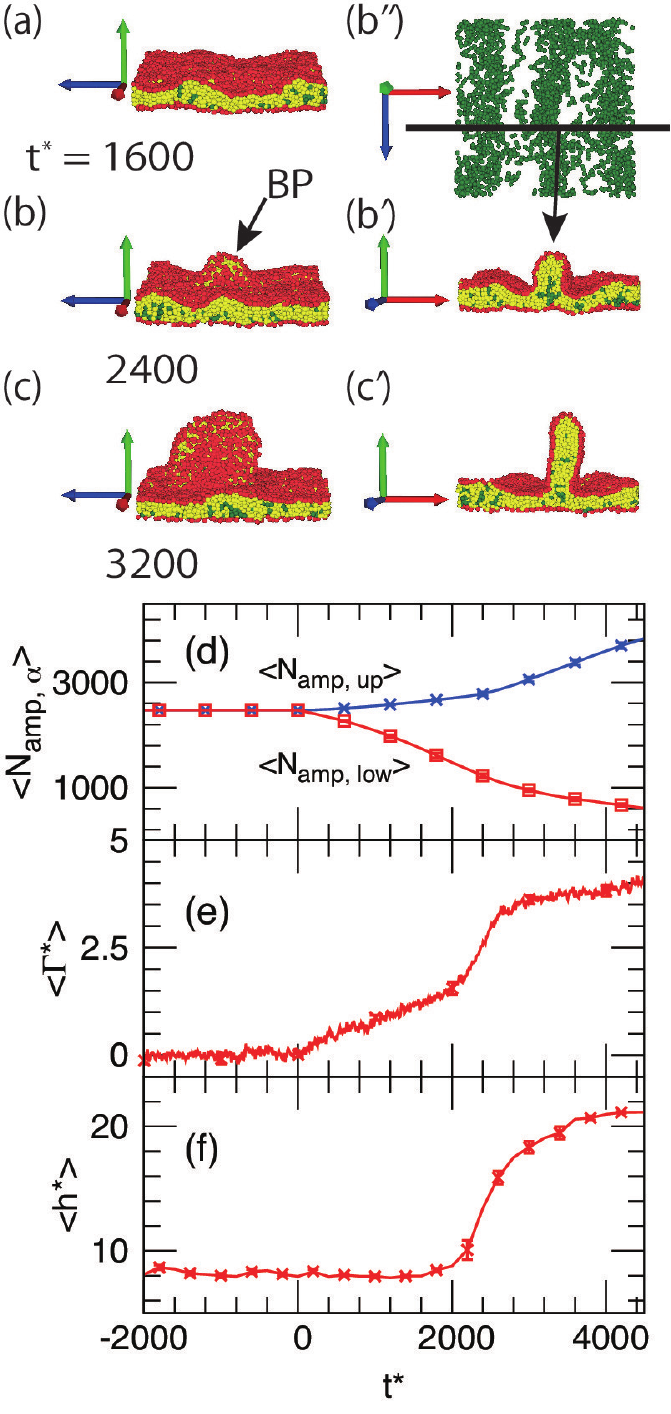}
 \caption{
   (a--c') Sequential snapshots of a bilayer sheet protrusion~(BP) from a flat
   membrane at $p^{*}_{\mathrm{f}}=20$, $p^{*}_{\mathrm{r}}=180$,
   $\gamma^{*}_{\mathrm{sol}}=2$, and $\gamma^{*}_{\mathrm{TT}}=2$.
   The number presents the simulation time $t^{*}$.
   (a), (b), (c) Bird's-eye view. (b'), (c') Cross sections of (b), (c) in front view.
   (b'') Only EOs of (b) are shown from the $z$ direction. Bold line of (b'')
   represents the cross section of (b').
   Time development of (d) $\langle N_{\mathrm{amp,up}}\rangle$ and
   $\langle N_{\mathrm{amp,low}}\rangle$, (e) surface tension $\Gamma$ , and
   (f) maximum height $h$ of the membrane.
   The error bars are calculated from eight independent runs.
   Symbols are shown for several data points.
   Smoothed data are shown for $\Gamma$.
 }
 \label{fig:BPform}
\end{figure}

First, we consider the shape transformations of the flat membrane by
chemical reactions~(see Fig.~\ref{fig:BPform}).
The HS concentration difference causes 
different time developments of the number of amphiphilic molecules between the upper and lower
leaflets as shown in Fig.~\ref{fig:BPform}(d): the number of amphiphilic molecules
$N_{\mathrm{amp}, \mathrm{up}}$ ($N_{\mathrm{amp}, \mathrm{low}}$) of
the upper (lower) leaflet increases (decreases).
On the lower leaflet, the condensation reaction is very slow owing to the low HS
concentration of the lower solution, with the result that the hydrolysis reaction largely proceeds.
The EOs are then produced at the lower leaflet and
embedded in the bilayer membrane, as indicated by the green color in
Fig.~\ref{fig:BPform}(a).
On the other hand, the condensation reaction largely proceeds on
the upper leaflet owing to the high HS concentration of the upper solution,
with the result that amphiphilic molecules are synthesized in the upper leaflet.

The resulting asymmetric distribution of amphiphilic molecules produces a negative surface
tension in the upper leaflet and a positive surface tension in the lower
leaflet, so that the upper (lower) leaflet is compressed (expanded).
This compression induces a large  undulation of the upper leaflet.
Figures~\ref{fig:BPform}(e) and (f) show the time development
of the surface tension $\Gamma$ of the bilayer and
maximum height $h$ of the membrane.
The surface tension (mechanical frame tension) is estimated as
$\Gamma = (P_{zz} - (P_{xx} + P_{yy})/2) L_{z}$
from the stress tensor~\cite{venturoli2006mesoscopic}.
Note that $\Gamma$ is the sum of two surface tensions: the surface tension
  $\Gamma_{\mathrm{up}}$ of the upper leaflet and the surface tension
  $\Gamma_{\mathrm{low}}$ of the lower leaflet.
  Since the hydrolysis reaction is faster than
   the condensation reaction,
 $\langle N_{\mathrm{amp,low}}\rangle$ has a greater slope than $\langle N_{\mathrm{amp,up}}\rangle$
, and the total surface tension $\Gamma$ then increases 
(see the data at $0<t^{*}< 2000$ in Figs.~\ref{fig:BPform}(d) and (e)).
A further increase in the surface tension induces the
buckling of the upper leaflet into the protrusion of a bilayer sheet
(indicated by arrows in Fig.~\ref{fig:BPform}(b)).
We hereafter refer to this deformation as bilayer protrusion (BP) formation.
The edge of the BP is tongue-shaped owing to the edge line tension.
Because the line tension of the branching junction between the BP and the bilayer is
low in the high-EO-density area, the
BP grows in the high-EO-density area as shown in Figs.~\ref{fig:BPform}(b') and
(b'').
The surface tension $\Gamma$  and maximum height $h$ of the membrane
rapidly increase
during BP formation at $t^{*} \simeq 2000$ (see Figs.~\ref{fig:BPform}(e) and (f)).
The BP releases the compressive (negative) surface tension $\Gamma_{\mathrm{up}}$ in the upper leaflet,
which increases $\Gamma$.

\begin{figure}[htbp]
  \centering
  \includegraphics[width=70mm]{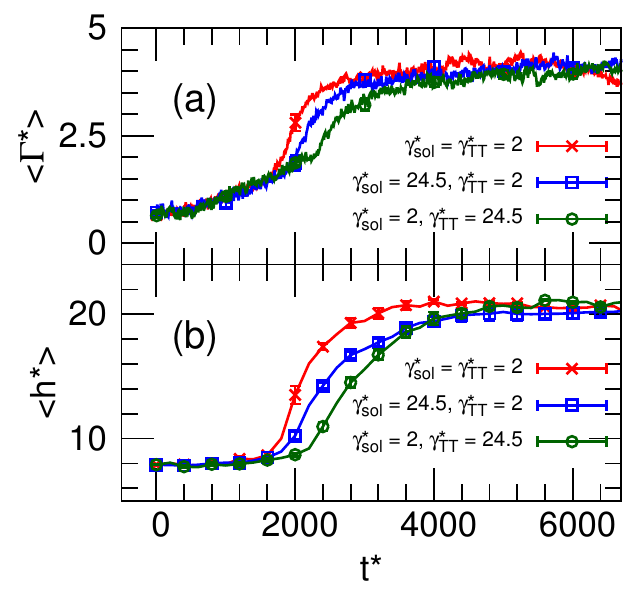}
  \caption{Time development of (a) the surface tension $\Gamma$ and (b) maximum height $h$
    of membrane for different viscosities.
    The error bars are calculated from eight independent runs.
   Symbols are shown for several data points.
    Smoothed data are shown for $\Gamma$.
  }
  \label{fig:stress_shape_gamma}
\end{figure}

  To investigate the effects of the viscosities of the surrounding fluids and
  membrane on the stress
  relaxation timescales of BP formation,
  we started simulations with different $\gamma_{\mathrm{sol}}$ and
  $\gamma_{\mathrm{TT}}$ values
from membranes equilibrated at the EO ratio $N_{\mathrm{EO}}/N_{\mathrm{amp}}
\simeq 0.06$ by stopping the chemical reactions.
When the viscosity $\eta_{\mathrm{sol}}$ of the surrounding fluids 
and the effective membrane viscosity  $\eta_{\mathrm{mb}}$ are increased roughly fivefold by changing
$\gamma_{\mathrm{sol}}$ and $\gamma_{\mathrm{TT}}$, 
BP formation is delayed by $400\tau$ and $800\tau$, respectively 
(see Fig.~\ref{fig:stress_shape_gamma}).
  This larger delay shows that the viscosity in the membrane has a
  stronger influence on BP formation.

\begin{figure}[htbp]
 \centering
 \includegraphics[width=70mm]{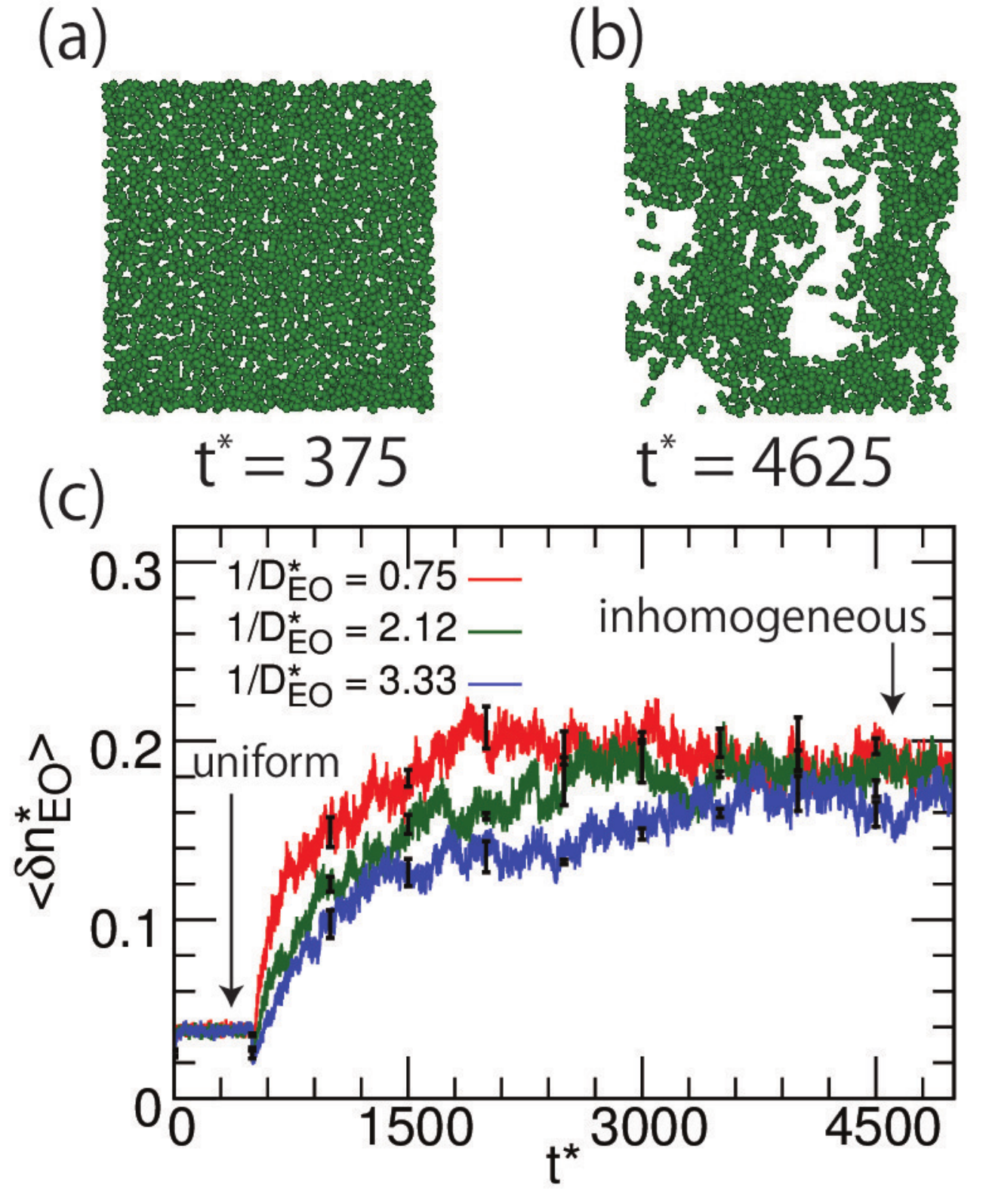}
 \caption{Time development of $\delta n_{\mathrm{EO}}$ of flat membrane for
   different $D_{\mathrm{EO}}$ values.
   Position restraints of EOs are removed at $t=500\tau$.
   Two snapshots represent only EOs in the bilayer membrane for $\gamma^{*}_{\mathrm{TT}}=2$.
   The left snapshot shows that EOs are uniformly distributed in the bilayer membrane.
   The right snapshot shows the inhomogeneous distribution of EOs.
   The error bars are calculated from three independent runs.
   Symbols are shown for several data points.
 }
 \label{fig:rho_inhomo_uni}
\end{figure}

In our simulation, the EOs are inhomogeneously distributed in the membrane.
To clarify whether this EO inhomogeneity is generated  thermodynamically or kinetically,
we simulate the equilibrium flat membrane in the absence of chemical reactions.
Figure~\ref{fig:rho_inhomo_uni}
shows the time development of the surface density inhomogeneity $\delta
n_{\mathrm{EO}}$ of the EOs.
The distribution of the EOs becomes inhomogeneous even if their initial distribution
is uniform. The EOs self-assemble into several clusters in the bilayer membrane (see
Fig.~\ref{fig:rho_inhomo_uni}(b)).
Thus, this inhomogeneity exists in thermal equilibrium.
The relaxation time $\tau_{\mathrm{inhomo}}$ from uniform to nonuniform spatial EO
distributions depends on the diffusion constant $D_{\mathrm{EO}}$ of the EOs.
Under faster EO diffusion,
$\delta n_{\mathrm{EO}}$ reaches equilibrium more rapidly (see Fig.~\ref{fig:rho_inhomo_uni}(c)).
This inhomogeneous nature is related to the orientational order of the bilayer
membranes~\cite{nakagawa2015morphological}. 
The orientational order of amphiphilic molecules is disturbed by the contacted EOs.

\subsection{Morphological changes at $v \simeq 1$}\label{sec:high_rv}
\begin{figure}[htbp]
 \centering
 \includegraphics[width=70mm]{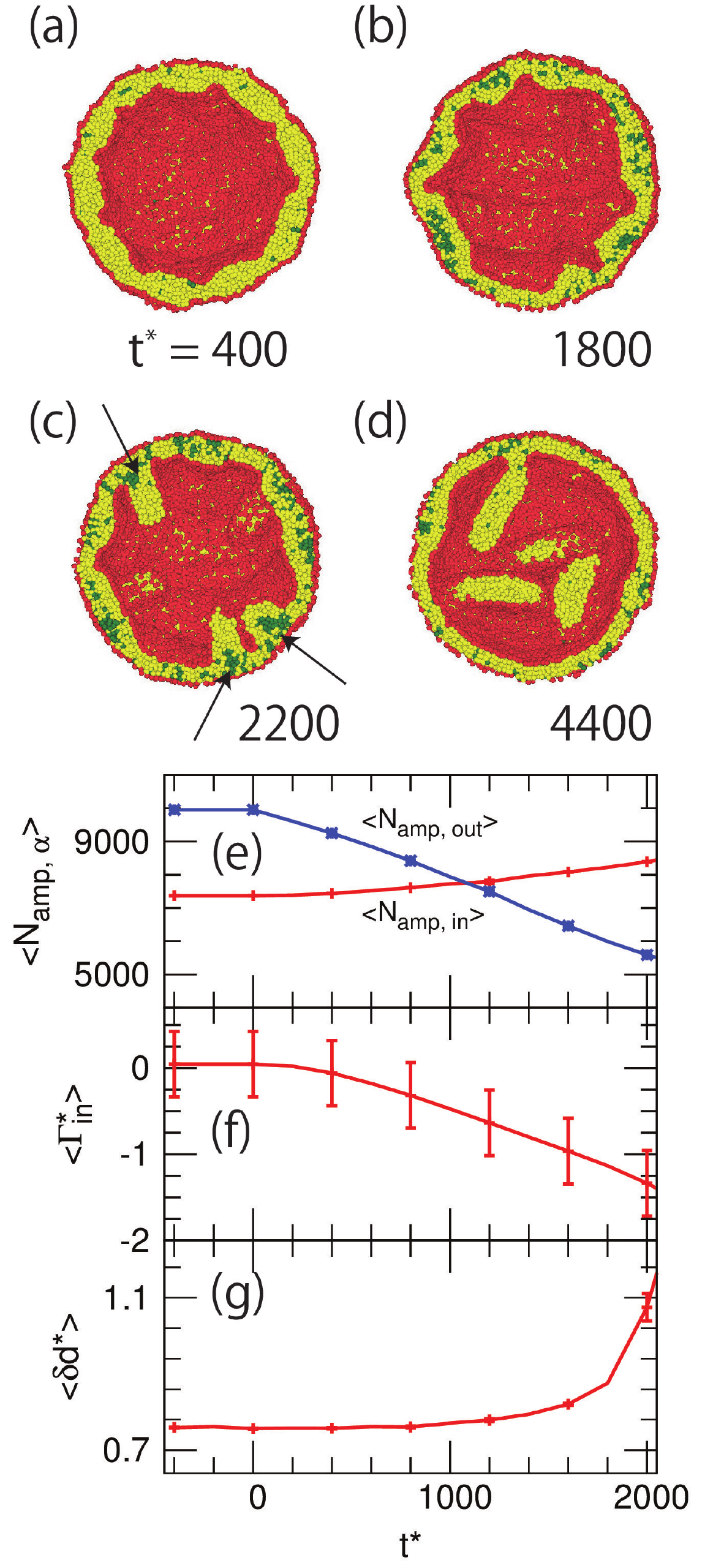}
 \caption{(a--d) Sequential snapshots of vesicle at $v \simeq 1$,
   $p^{*}_{\mathrm{f}} = 80$, $p^{*}_{\mathrm{r}} = 160$,
   $c_{\mathrm{in}}^{*} = 3$,
   $\gamma^{*}_{\mathrm{sol}} = 2$,
   and $\gamma^{*}_{\mathrm{TT}} = 18$.
   The numbers present the simulation time $t^{*}$.
The front halves of the vesicles are not displayed 
to show the inner structures of the vesicle.
   (e--g) Time development of (e) $\langle N_{\mathrm{amp},\mathrm{in}}
   \rangle$ and $\langle N_{\mathrm{amp},\mathrm{out}} \rangle$,
   (f) surface tension $\langle \Gamma_{\mathrm{in}} \rangle$ of inside
   monolayer calculated by eqn~(\ref{eq:surf}), and (g) thickness inhomogeneity
   $\langle \delta d \rangle$.
   The error bars are calculated from eight independent runs.
   Symbols are shown for several data points.
 }
 \label{fig:col_highrvol}
\end{figure}

Next, we consider the shape transformations of a vesicle at $v \simeq 1$ and 
$c^{*}_{\mathrm{in}} = 3$ (see Fig.~\ref{fig:col_highrvol}).
This concentration difference causes the transport of amphiphilic molecules 
from the outer leaflet to the inner leaflet as in the flat membrane (see Fig.~\ref{fig:col_highrvol}(e)).

We estimate the surface tension
$\Gamma_{\mathrm{in}}$ of the inner leaflet in the following manner.
Since vesicles before shape transformations are nearly spherical because $v \simeq 1$,
the surface tension $\Gamma_{\mathrm{in}}$ of the inner leaflet
is approximately estimated as
 \begin{align}
   \Gamma_{\mathrm{in}} = K_{A} \dfrac{4\pi(R_{\mathrm{ves}} - h_{\mathrm{neut}})^{2} - N_{\mathrm{amp},\mathrm{in}} a_{0}}{N_{\mathrm{amp},\mathrm{in}} a_{0}},
   \label{eq:surf}
 \end{align}
where $a_{0} = 0.52r_{\mathrm{cut}}^2$ is the area per lipid molecule in the tensionless membrane,
$R_{\mathrm{ves}}$ is the radius of the vesicle, and
$h_{\mathrm{neut}} = 0.9r_{\mathrm{cut}}$ is the distance between the inner leaflet
neutral surface and the bilayer mid-plane.
The negative surface tension of $\Gamma_{\mathrm{in}}$ induces the
buckling of the inner leaflet, leading to the formation of multiple BPs
(indicated by arrows in Fig.~\ref{fig:col_highrvol}(c)).
This BP formation process can be captured by the time development of the
standard deviation of the bilayer
thickness $\delta d$ (the calculation method is described in Appendix \ref{apdx:calc_thick}) 
as shown in Fig.~\ref{fig:col_highrvol}(g). During the undulation,
$\delta d$ gradually increases, and after the buckling starts, $\delta d$ rapidly increases at $t^{*}>1500$.

\begin{figure}[htbp]
   \centering
   \includegraphics[width=70mm]{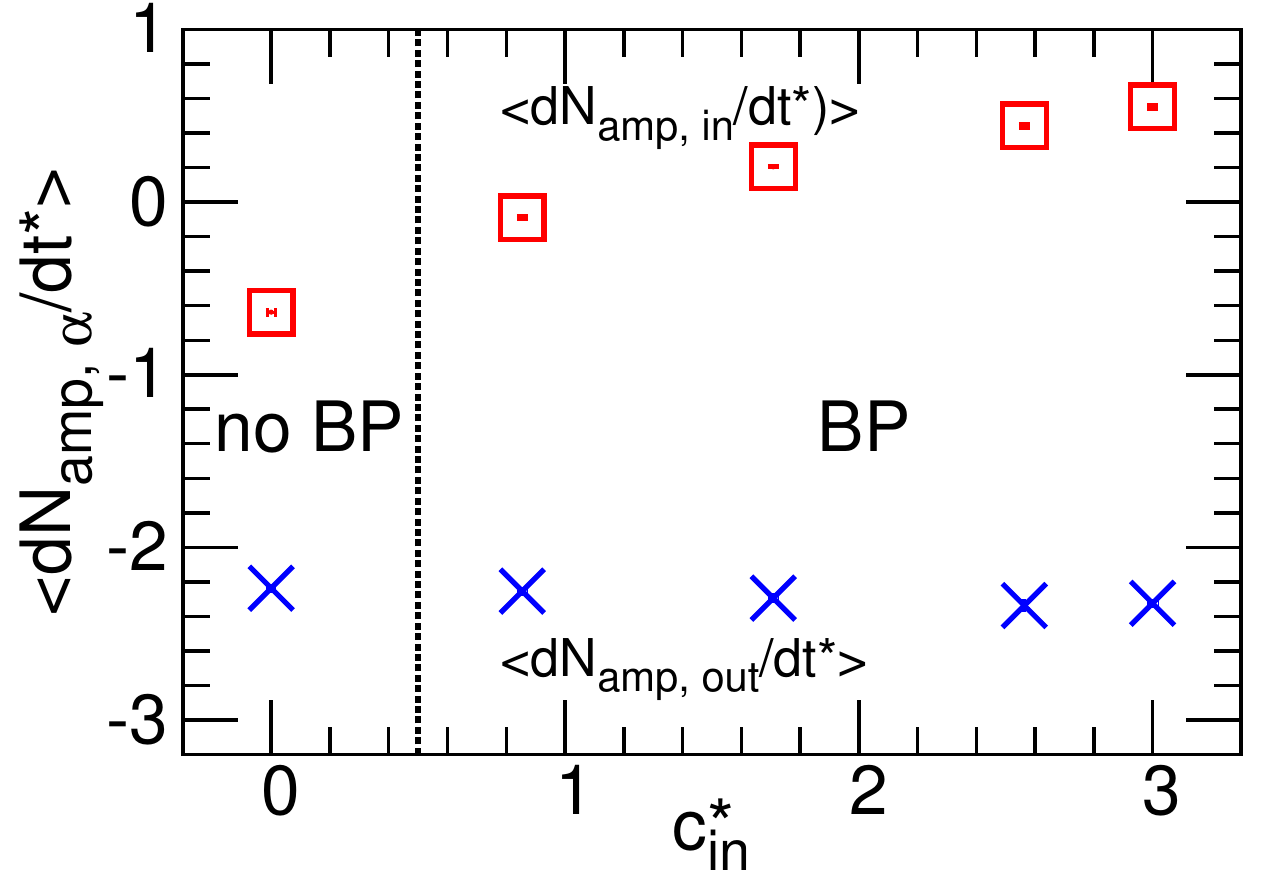}
   \caption{Reaction rates of amphiphilic molecules in each leaflet as a
     function
     of the initial concentration $c^{*}_{\mathrm{in}}$
       inside the vesicle at 
     $\gamma^{*}_{\mathrm{sol}} = \gamma^{*}_{\mathrm{TT}}=2$,
     $p^{*}_{\mathrm{f}}=80$, and $p^{*}_{\mathrm{r}}=160$.
     BPs form at $c^{*}_{\mathrm{in}} = 0.854, 1.71, 2.56$, and $3$. 
     The vertical dotted line serves as a guide for the eye for the threshold of BP formation.
   }
   \label{fig:chem_vs_c}
  \end{figure}

BP formation depends on the initial HS concentration
$c_{\mathrm{in}}$ inside the vesicle as
shown in Fig.~\ref{fig:chem_vs_c}.
As $c_{\mathrm{in}}$ increases, 
the reaction rate $dN_{\mathrm{amp}, \mathrm{in}}/dt$ increases, whereas
$dN_{\mathrm{amp}, \mathrm{out}}/dt$ shows little dependence.
BP formation becomes faster with increasing $c_{\mathrm{in}}$ as a result of the
increase in $dN_{\mathrm{amp}, \mathrm{in}}/dt$.
BP formation occurs everywhere except at $c^{*}_{\mathrm{in}} = 0$.
Thus, the synthesis of amphiphilic molecules
and the resulting negative surface tension of the inner leaflet are necessary for BP formation.

\begin{figure}[htbp]
  \centering
  \includegraphics[width=70mm]{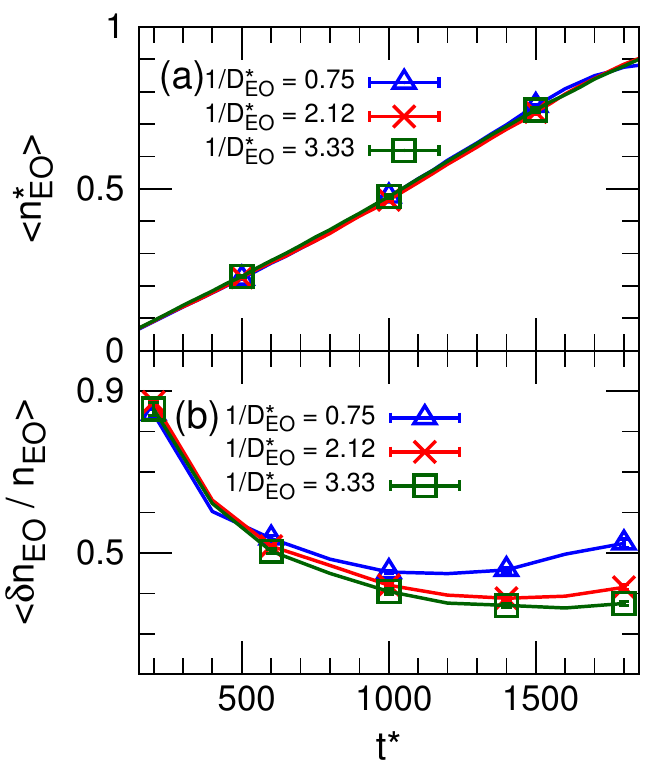}
  \caption{Time development of (a) the average $\langle n_{\mathrm{EO}} \rangle$
    and (b) standard deviation 
 $\langle \delta n_{\mathrm{EO}} / n_{\mathrm{EO}} \rangle$ of the surface density of EOs
    for various $D_{\mathrm{EO}}$ values at $p^{*}_{\mathrm f} = 80$,
    $p^{*}_{\mathrm r} = 160$, and  $\gamma^{*}_{\mathrm{sol}}=2$.
    $\gamma^{*}_{\mathrm{TT}}$ is varied from 2 to 24.5 to change $D_{\mathrm{EO}}$.
    The error bars are calculated from eight independent runs. \label{fig:rho_inhomo}
   Symbols are shown for several data points.
  }
  \end{figure}

As shown in the flat membrane simulation in Section~\ref{sec:flat}, the BPs
protrude from the EO clusters~(Figs.~\ref{fig:BPform} (b') and (b'')).
Thus, these clusters accelerate BP formation.
Hence, we next examine the spatial inhomogeneity of the EOs in the bilayer membrane
and its relation with BP formation.
The spatial inhomogeneity of the EOs is determined by two processes: the synthesis
and diffusion of the EOs. The synthesis of the EOs occurs uniformly
on the outer leaflet, leading to a homogeneous EO distribution. The
characteristic timescale $\tau_{\mathrm{homo}}$ of this process is the
reciprocal of the EO synthesis speed. On the other hand, the EOs in the bilayer
membrane self-assemble into several clusters, and thus the spatial distribution
of the EOs becomes inhomogeneous as discussed in the case of the flat membrane. This characteristic timescale
$\tau_{\mathrm{inhomo}}$ of the EO assembly into clusters decreases
(increases) for faster (slower) diffusion of the EOs.
The relationship between these two timescales $\tau_{\mathrm{homo}}$ and
$\tau_{\mathrm{inhomo}}$ affects the inhomogeneity of the EOs.
When $\tau_{\mathrm{homo}} > \tau_{\mathrm{inhomo}}$, the EOs self-assemble into
clusters rapidly, but otherwise they remain close to the uniform distribution.
We confirm this tendency by simulations in which
$\tau_{\mathrm{inhomo}}$ is varied by changing the diffusion constant
$D_{\mathrm{EO}}$ of the EOs.
Figures~\ref{fig:rho_inhomo}(a) and (b) show the time development of the mean
surface density of the EOs $n_{\mathrm{EO}}$ and its inhomogeneity $\delta
n_{\mathrm{EO}}(t)$, which is defined as the standard deviation of the surface
density of the EOs, for different $D_{\mathrm{EO}}$ values.
The EO density is calculated by using a bin of solid
angle $\omega = 16a_{0}/(4 \pi R^{2}_{\mathrm{ves}})$ in the spherical vesicle.
The time development of $n_{\mathrm{EO}}(t)$ is not affected
by the change in $D_{\mathrm{EO}}$, as shown in Fig.~\ref{fig:rho_inhomo}(a),
so that $\tau_{\mathrm{homo}}$ does not depend on $D_{\mathrm{EO}}$.
On the other hand, the time development of $\delta n_{\mathrm{EO}}$ clearly depends on
$D_{\mathrm{EO}}$ at the late stage $t^{*} \in [1000, 1800]$.
In the beginning of the chemical reaction (at $t^{*} < 1000$ in
Fig.~\ref{fig:rho_inhomo}(b)), the spatial inhomogeneity, $\delta n_{\mathrm{EO}}$,
decreases as the EOs are synthesized.
For fast diffusion $1/D^{*}_{\mathrm{EO}} = 0.75$, the
EOs form clusters, and $\delta n_{\mathrm{EO}}$
increases at $t^{*} \in [1000, 1800]$.
Thus, $\tau_{\mathrm{homo}} > \tau_{\mathrm{inhomo}}$ is satisfied.
However, for slow diffusion, $1/D^{*}_{\mathrm{EO}} = 2.12$ and $3.33$,
the EOs are uniformly distributed even at $t^{*} \in [1000, 1800]$,
and the cluster formation occurs at later stages.

\begin{figure}
 \centering
 \includegraphics[width=70mm]{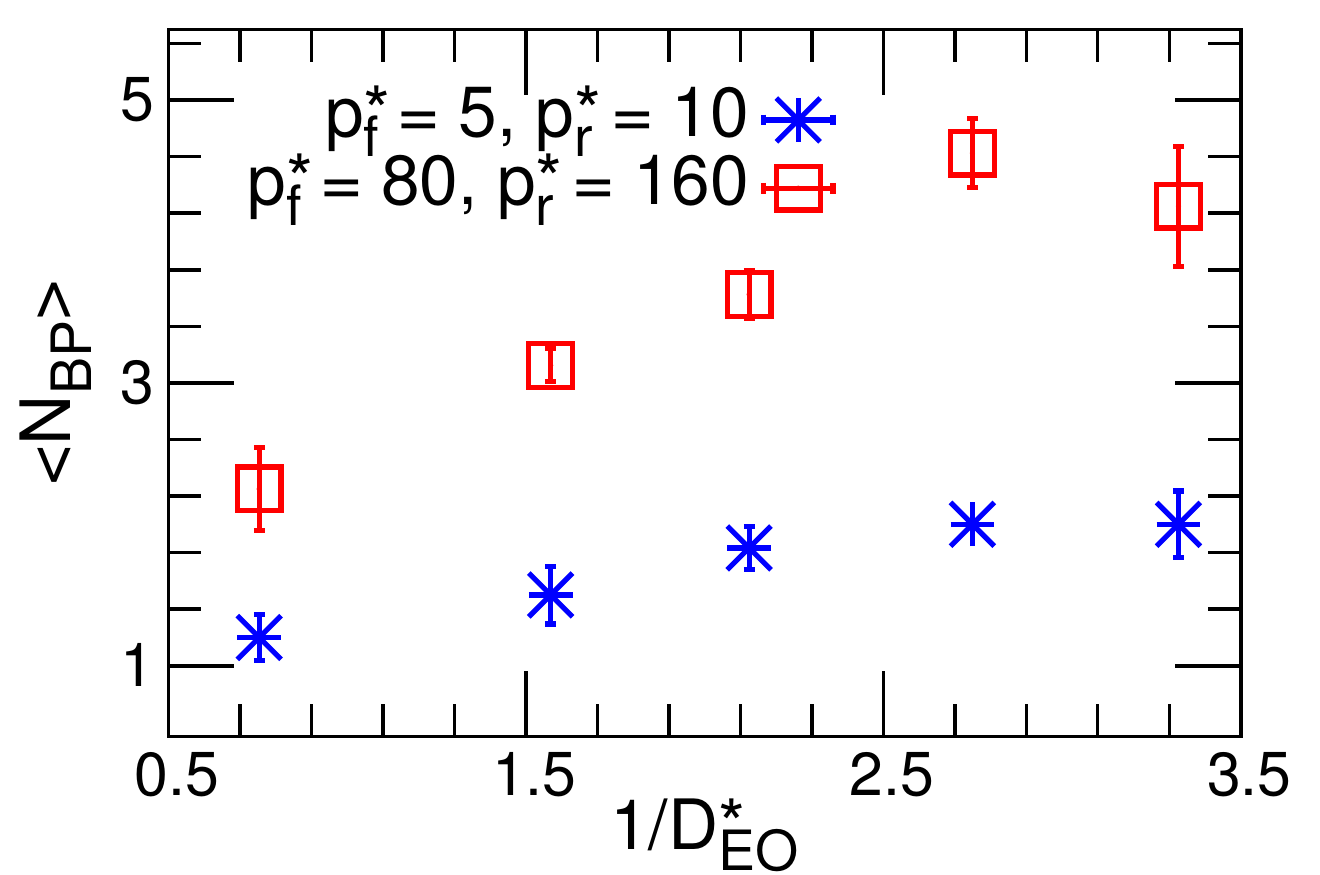}
 \caption{Number $N_{\mathrm{BP}}$ of BPs as a function of $1/D_{\mathrm{EO}}$
   for two reaction rate settings of $p_{\mathrm{f}}$ and $p_{\mathrm{r}}$ at 
$\gamma^{*}_{\mathrm{sol}}=2$.
   The error bars are calculated from six independent runs.
 }
 \label{fig:protru}
\end{figure}

More BPs form at slower diffusion and faster reactions as shown in Fig.~\ref{fig:protru}.
At small values of the diffusion constant $D_{\mathrm{EO}}$,
slower EO cluster formation delays BP formation.
Since the synthesis rate of EOs does not depend on the
  $D_{\mathrm{EO}}$ as shown in Fig.~\ref{fig:rho_inhomo} (a),
  the number of EOs at the BP formation increases at
  lower $D_{\mathrm{EO}}$. At high reaction rates $p^{*}_{\mathrm f}$ and
$p^{*}_{\mathrm r}$,
the surface tension decreases more rapidly.
In both cases, at BP formation,
  the larger number of EO clusters exist in the bilayer membrane, and
the membrane is under a greater compressive tension.
This leads to the formation of more BPs at the same time instead of a single large BP.

\begin{figure}
 \centering
 \includegraphics[width=70mm]{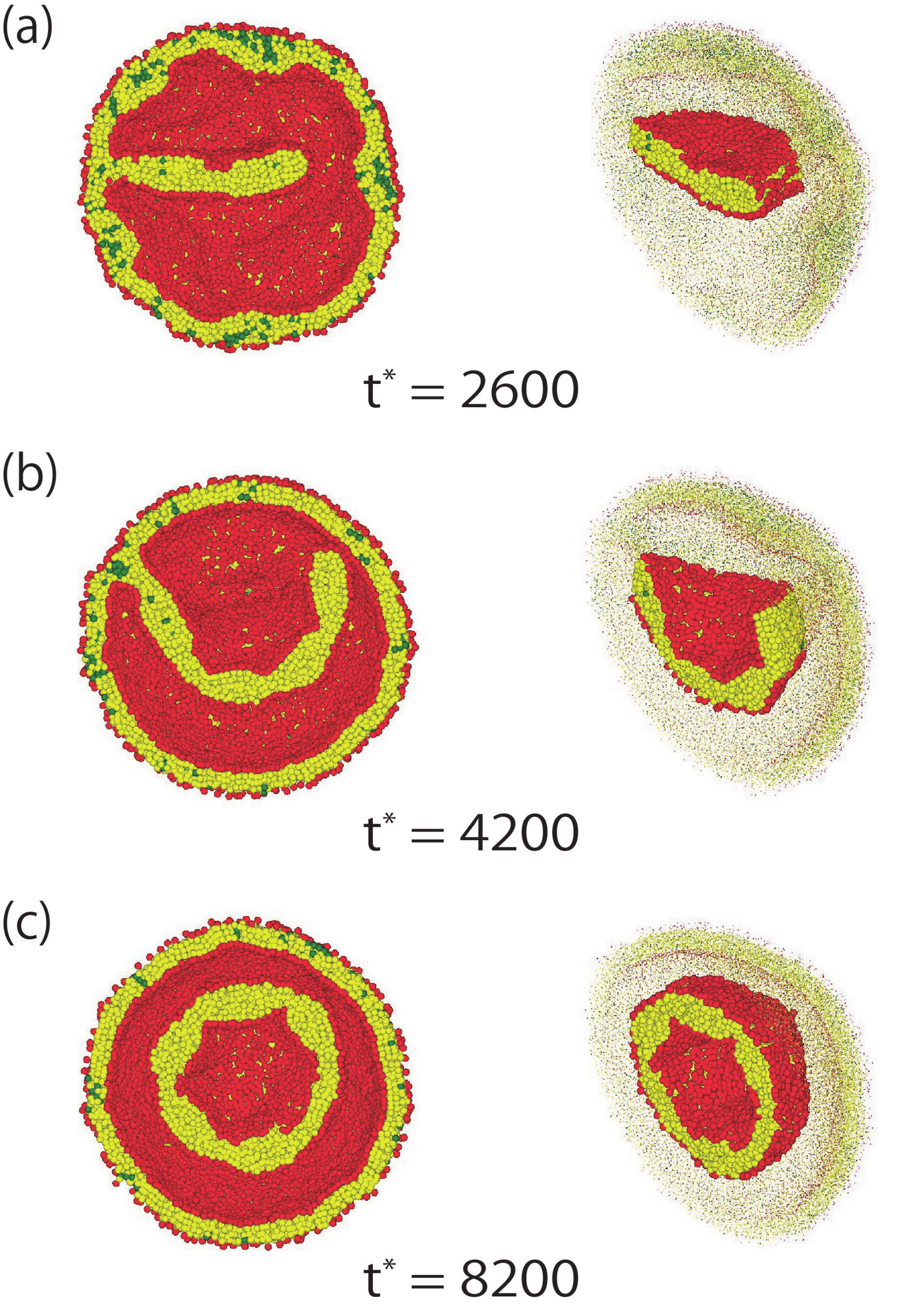}
 \caption{
 (a--c) Sequential snapshots of shape transformation from the flat BP
   disk to the vesicle at $v \simeq 1$,
   $p^{*}_{\mathrm{f}} = 80$, $p^{*}_{\mathrm{f}} = 160$,
   $c_{\mathrm{in}}^{*}=3$, $\gamma^{*}_{\mathrm{sol}}=2$, and
   $\gamma^{*}_{\mathrm{TT}}=2$. Cross-sectional images are shown.
   In the right panels, the membrane inside the vesicles are extracted from the
   left panels. The number presents the simulation time~$t^{*}$.
 }
 \label{fig:bp_to_ves}
\end{figure}

  After the BP formation,
  the BP bends and subsequently
  transforms into a spherical vesicle as shown in Fig.~\ref{fig:bp_to_ves} for small number $N_{\mathrm{BP}}$ of BPs.
  Since the initial shape of the BP is a nearly flat disk,
  this shape transformation can be understood as the shape
  transformation from a flat bilayer disk to a spherical vesicle.
  This type of shape transformation is described by
  the theory by Fromherz~\cite{fromherz1983lipid}.
  Large flat bilayer disks are energetically unstable at
$A_{\mathrm{mb}}>16\pi[(2\kappa + \bar{\kappa})/\Gamma_{\mathrm{edge}} ]^2$,
where $\Gamma_{\mathrm{edge}}$, $A_{\mathrm{mb}}$, and $\bar{\kappa}$ are
  the edge line tension, area, and saddle-splay modulus of the
  membrane, respectively. 
  Therefore, the sufficiently grown flat BP spontaneously transforms into the spherical vesicle.
  At a large number $N_{\mathrm{BP}}$ of BPs, the contacts between the BPs prevent
  the shape transformation to the spherical shapes; thus the BPs remain flat.

\subsection{Morphological changes at $v \simeq 0.5$}\label{sec:low_rv}
\begin{figure}[htbp]
 \centering
 \includegraphics[width=70mm]{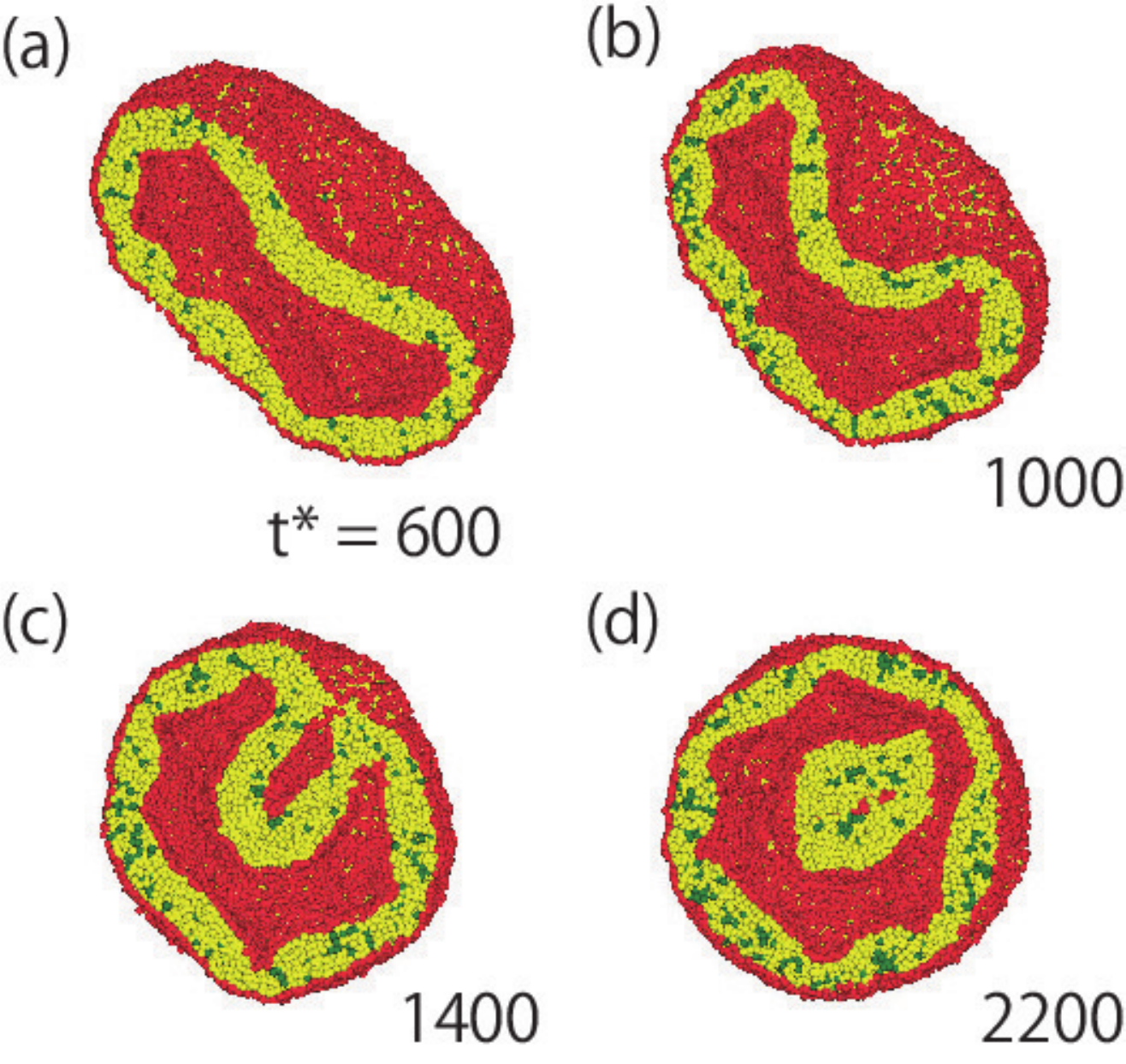}
 \caption{Sequential snapshots of bud formation at
   $\gamma^{*}_{\mathrm{sol}}  = \gamma^{*}_{\mathrm{TT}} = 2$,
   $p^{*}_{\mathrm f} = 20$, $p^{*}_{\mathrm r} = 180$, and $c_{\mathrm{in}}^{*} = 3$.
   Cross-sectional images are shown.
 }
 \label{fig:budding_low}
\end{figure}

Next, we consider the shape transformations at a low reduced volume, $v \simeq 0.5$.
The initial shape is discocyte (see Fig.~\ref{fig:budding_low}(a)).
The same settings as in Section~\ref{sec:high_rv} are considered: 
the HS concentration is different inside and outside the vesicle.
Thus, the hydrolysis and condensation reactions mainly occur on the outer and
inner leaflets, respectively.

For the low reduced volume, large low-wavelength fluctuations of the bilayer
are permitted unlike for $v \simeq 1$, leading to a different type of shape transformation, 
budding into a stomatocyte.
Figure~\ref{fig:budding_low} shows the typical shape transformation.
 As the hydrolysis and condensation reactions
proceed, a dimple invagination forms (Fig.~\ref{fig:budding_low}(b)), and eventually
the discocyte transforms into a stomatocyte (see Figs.~\ref{fig:budding_low}(c) and (d)).
A decrease and increase in the amphiphilic molecular densities of the outer and inner leaflets, respectively,
 cause an effective negative spontaneous
 curvature according to the ADE model so that the inner bud is stabilized.
 After the budding,
a further increase of amphiphilic molecular density of the inner leaflet causes BP
formation, as in the $v\simeq 1$ case.

\begin{figure}[htbp]
 \centering
 \includegraphics[width=70mm]{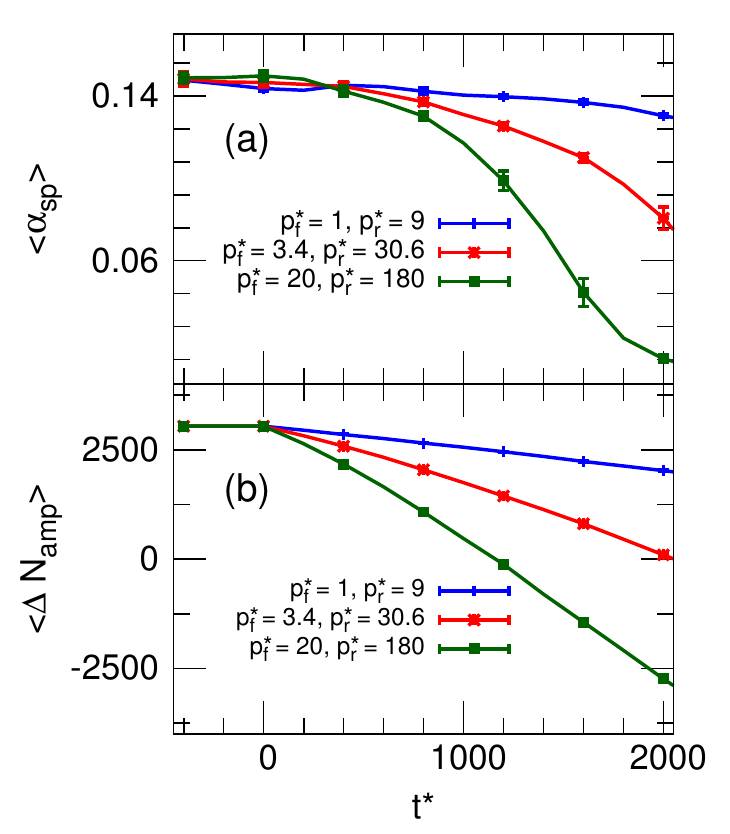}
 \caption{Time development of (a) asphericity $\langle \alpha_{\mathrm{sp}} \rangle$,
   (b) amphiphilic molecular number difference $\langle \Delta N_{\mathrm{amp}} \rangle$ between the inner
   and outer leaflets at $c_{\mathrm{in}}^{*} = 3$,
   $\gamma^{*}_{\mathrm{sol}} = 24.5$, and
   $\gamma^{*}_{\mathrm{TT}} = 2$.
   The error bars are calculated from eight independent runs.
   Symbols are shown for several data points.
   \label{fig:ade_emb_asp}}
\end{figure}

In order to clarify the relationship between
shape transformations and chemical reactions, we calculate the time development of the
asphericity $\alpha_{\mathrm{sp}}$~\cite{theodorou1985shape}
(Fig.~\ref{fig:ade_emb_asp}(a)), and the
amphiphilic molecular number difference, $\Delta N_{\mathrm{amp}} =
N_{\mathrm{amp, out}} - N_{\mathrm{amp, in}}$ between the inner and outer leaflets
(Fig.~\ref{fig:ade_emb_asp}(b)).
The asphericity is the degree of deviation from a spherical 
shape and is defined as $\alpha_{\rm {sp}} = [(\lambda_1-\lambda_2)^2 + 
  (\lambda_2-\lambda_3)^2+(\lambda_3-\lambda_1)^2]/2(\lambda_1+\lambda_2+\lambda_3)^2$,
where $\lambda_1$, $\lambda_2$, and $\lambda_3$ are the 
eigenvalues of the gyration tensor of the vesicle.
It can distinguish the stomatocyte ($\alpha_{\mathrm{sp}}\simeq 0$) 
and the discocyte ($0.15 \lesssim \alpha_{\mathrm{sp}} \lesssim 0.25$)~\cite{nogu05}.
When the chemical reaction process speeds up by increasing $p_{\mathrm{f}}$
and $p_{\mathrm{r}}$,
the shape transformation also speeds up (from $p^{*}_{\mathrm{f}}=1,
p^{*}_{\mathrm{r}}=9$ to $p^{*}_{\mathrm{f}}=20,
p^{*}_{\mathrm{r}}=180$ in Fig.~\ref{fig:ade_emb_asp}(a)).
As the chemical reactions proceed,
 $\Delta N_{\mathrm{amp}}$ linearly decreases in time, but
$\alpha_{\mathrm{sp}}$ nonlinearly decreases.
This rapid change corresponds to the shape transformation from the discocyte to the stomatocyte.

\begin{figure}[htbp]
 \centering
 \includegraphics[width=60mm]{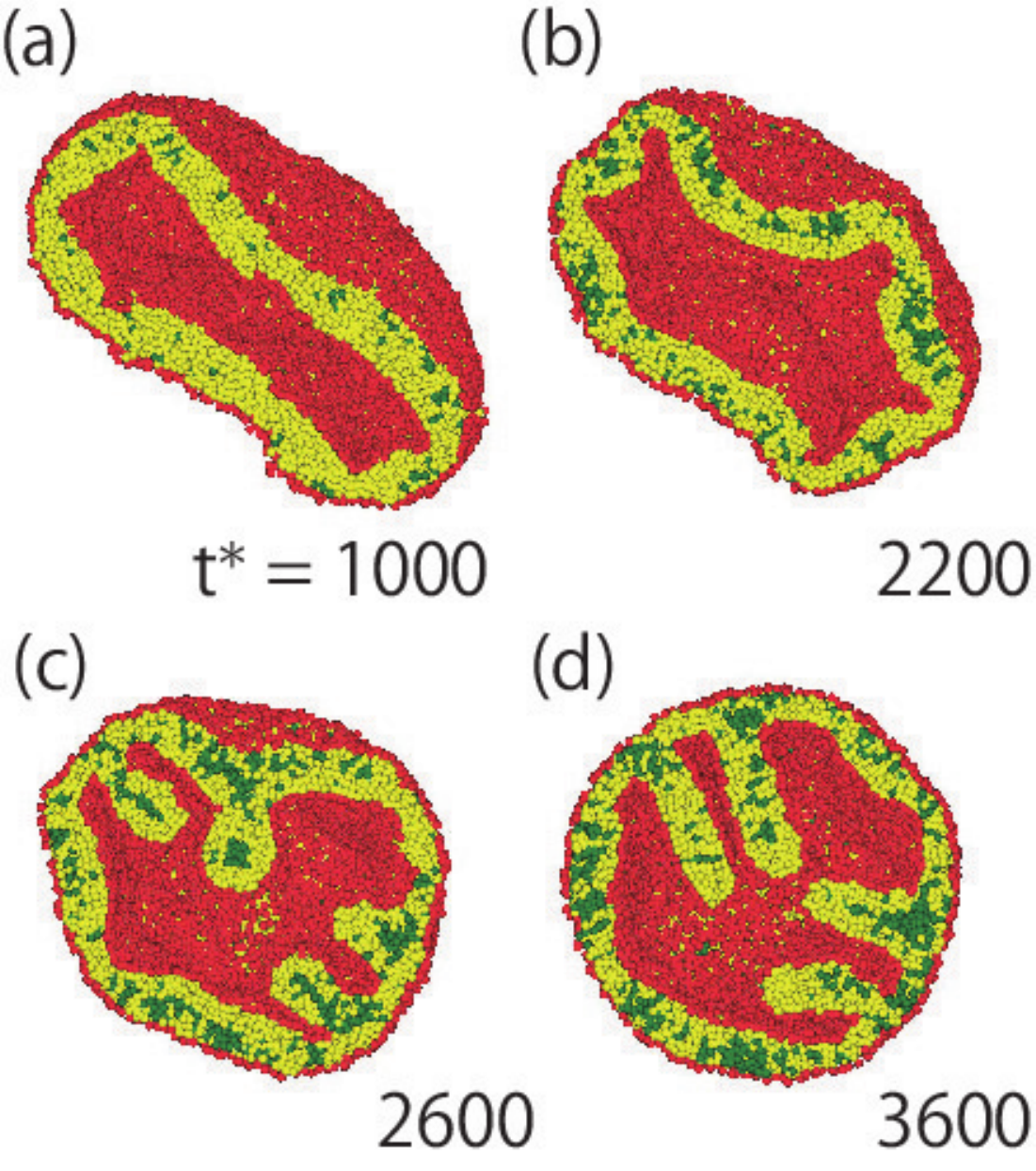}
 \caption{Sequential snapshots of BP formation without budding at
   $\gamma^{*}_{\mathrm{sol}} = 24.5$, $\gamma^{*}_{\mathrm{TT}} = 2$,
   $p^{*}_{\mathrm f} = 5$, $p^{*}_{\mathrm r} = 45$, and $c_{\mathrm{in}}^{*} = 3$.
   Cross-sectional images are shown.
 }
 \label{fig:collapse_low}
\end{figure}

Interestingly, BP formation without budding occurs when the viscosity
$\eta_{\mathrm{sol}}$ is increased (see Fig.~\ref{fig:collapse_low}).
Initially, the bilayer bends inward (Fig.~\ref{fig:collapse_low}(b)) as in the budding, 
but these invaginations transform into BPs (Figs.~\ref{fig:collapse_low}(c) and (d)).

\begin{figure}[htbp]
 \centering
 \includegraphics[width=70mm]{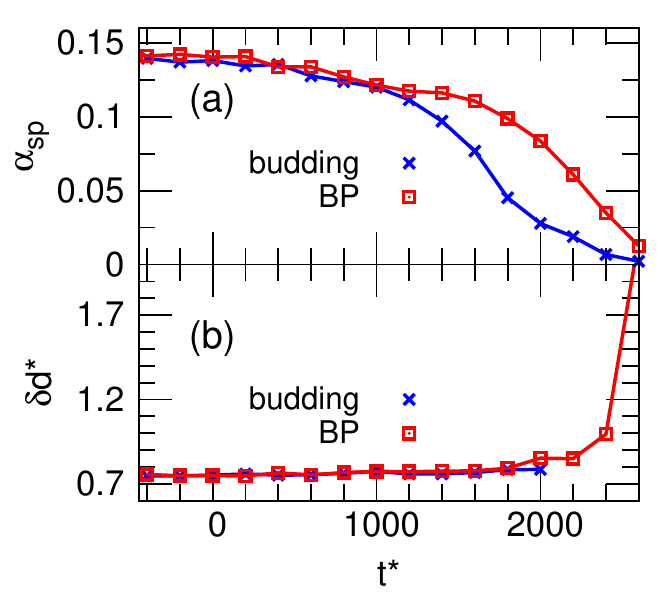}
 \caption{Time development of (a) asphericity $\alpha_{\mathrm{sp}}$ and (b) bilayer thickness
   inhomogeneity $\delta d$ at $c_{\mathrm{in}}^{*} = 3$,
   $p^{*}_{\mathrm f} = 5$, $p^{*}_{\mathrm r} = 45$,
   $\gamma^{*}_{\mathrm{sol}} = 24.5$, and
   $\gamma^{*}_{\mathrm{TT}} = 2$.
   Blue and red lines represent budding and BP formation without budding,
   respectively.
   Symbols are shown for several data points.
   \label{fig:asp_thick}
 }
\end{figure}

In order to distinguish the two types of shape transformations (budding (Fig.~\ref{fig:budding_low}) and
BP formation without budding (Fig.~\ref{fig:collapse_low})), we calculate the bilayer thickness inhomogeneous~$\delta d$
during shape transformation from $\alpha_{\mathrm{sp}} \simeq 0.14$ to $\alpha_{\mathrm{sp}} \simeq 0$ 
(see Fig.~\ref{fig:asp_thick}).
For BP formation, $\delta d$ is diverged when
$\alpha_{\mathrm{sp}}$ changes from 0.14 to 0.025. For bud formation, $\delta d$ is not diverged.
We set $\delta d_{\mathrm{th}} = r_{\mathrm{cut}}$ as the threshold value to
determine the divergence of the thickness inhomogeneity $\delta d$.
When $\delta d > \delta d_{\mathrm{th}}$
is satisfied from $\alpha_{\mathrm{sp}} \simeq 0.14$ to $\alpha_{\mathrm{sp}} \simeq 0$,
this shape transformation is regarded as BP formation.

\begin{figure}[htbp]
  \centering
  \includegraphics[width=60mm]{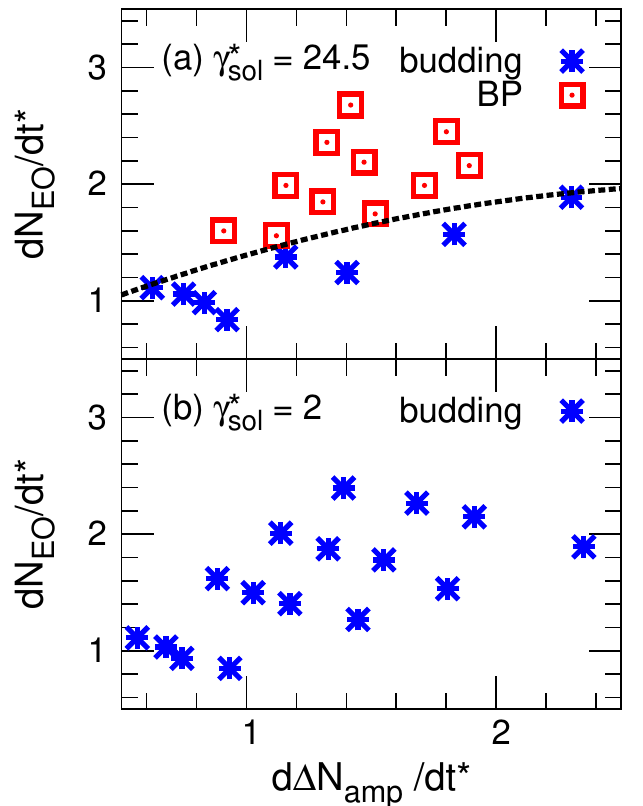}
  \caption{Dynamic phase diagram of shape transformations
    at (a) $\gamma^{*}_{\mathrm{sol}} = 24.5$ and
    (b) $\gamma^{*}_{\mathrm{sol}} = 2$ for $\gamma^{*}_{\mathrm{TT}} = 2$. \label{fig:colbud_phase}
  }
 \end{figure}

Using this threshold of $\delta d$, we construct the dynamic phase diagram of the shape
transformations as shown in Fig.~\ref{fig:colbud_phase} for different
$\eta_{\mathrm{sol}}$ values.
At each data point, we carry out eight independent
runs. If the number of BP transformations without budding is more than four at one point,
the shape transformation is regarded as BP formation without budding.

The shape transformation pathway depends on not only $d \Delta N_{\mathrm{amp}}/dt$ 
but also on $d N_{\mathrm{EO}}/dt$ as shown in Fig.~\ref{fig:colbud_phase}(a).
As mentioned in Section~\ref{sec:high_rv}, BP formation is strongly affected by the
spatial distribution of the EOs. If the EOs are not sufficiently synthesized, the
resulting shape transformations are budding (low $d N_{\mathrm{EO}}/dt$
in Fig.~\ref{fig:colbud_phase}(a)) because a few EOs do not form clusters in the
bilayer. When the EO synthesis rate increases, the EOs self-assemble into clusters in the
bilayer, so that BP formation occurs. Thus, EO synthesis dramatically affects the
resulting shape transformation.

However, at low viscosity $\eta_{\mathrm{sol}}$, BP without budding does not
occur, as shown in Fig.~\ref{fig:colbud_phase}(b).
In this case, the bud formation timescale is shorter than the
timescale $\tau_{\mathrm{inhomo}}$ of the EO cluster formation. 
The area compressive stress caused by the chemical
reactions is released via bud formation before BP formation starts.
As pointed out by Sens~\cite{sens2004dynamics}, the surrounding solution that
has high viscosity suppresses bud formation. This suppression of bud
formation enhances BP formation.
As mentioned in Section~\ref{sec:flat}, BP formation is more slowed
by the membrane viscosity.
Thus, compared to budding, BP formation more frequently occurs
at low membrane viscosity $\eta_{\mathrm{mb}}$ and high solution viscosity $\eta_{\mathrm{sol}}$.

\section{Discussions and Conclusion}\label{sec:conc}
In this paper, we have shown the membrane shape transformations
induced by hydrolysis and condensation reactions. We use the coarse-grained molecular
simulation technique in which the hydrolysis and condensation reactions are taken
into account explicitly by the bond dissociation and binding.
  The asymmetric chemical conditions, which are widely observed
  both {\it in vitro} and {\it in vivo}, cause the transport of
  amphiphiles between outer and inner leaflets leading to the BP formations.
The growth process of BPs strongly depends on the EO density inhomogeneity that
is determined by the competition between two different dynamics:
diffusion of EOs and synthesis of EOs.
  At the faster EO diffusion compared to the EO synthesis, fewer BPs form.
At a low reduced volume, budding transformation also occurs.
The shape transformation pathway is affected by the EO synthesis rate and the
shear viscosity of the surrounding solution. By increasing the solution
viscosity $\eta_{\mathrm{sol}}$ while keeping the membrane viscosity
$\eta_{\mathrm{mb}}$ constant,
bud formation is suppressed so that BP formation is enhanced.
In the budding, the membrane mainly moves normal to the membrane surface,
but sliding between two leaflets occurs in BP formation.
Thus, the viscosity of the surrounding fluids affects budding more than it does BP formation,
while the viscosity in the membrane affects BP formation more.

Since the BPs transform into small spherical vesicles,
it may be difficult to distinguish them from the buds formed by the budding process by optical microscopy.
However, the inside of the small vesicles is filled with the solution originally inside of the vesicles,
while the inside of the buds is filled with the outer solution.
Thus, it can be experimentally identified by labeling the inner or outer solution.

Similar shape transformations in BP formation are observed in the
Langmuir monolayer in both experiments~\cite{lu2002folding,lee2008collapse} and
simulations~\cite{baoukina2008molecular}.
The compression leads to a collapse of the Langmuir monolayer into a bilayer
sheet.  The formed bilayer sheet finally transforms into the spherical
vesicle~\cite{baoukina2008molecular,gopal2001morphology},
which is also similar to the shape transformation in our simulations.
Other experiments, in which bilayer vesicles are composed of SOPC (1-stearoyl-2-oleoyl-sn-glycero-3-phosphocholine), 
C16:0-SM (N-palmitoyl-sphingomyelin), and Bodipy-sphingomyelin (a fluorescent tracer), 
show the invagination of vesicles under the injection of the enzyme sphingomyelinase~\cite{holopainen2000vectorial}.
Ceramide, which is produced by the hydrolysis of SM, segregates into a domain in
the membrane
and causes the invagination of vesicle.
We expect that such a domain helps BP formation in the inner leaflet of the vesicles.

Our results show that the relative viscosity ratio of the membrane and surrounding fluids
is significant in determining budding or BP formation.
We believe that two competing shape transformations accompanied by  lateral and
normal membrane motions are generally controlled by the viscosity ratio;
the former and latter dynamics are slowed down by increases in viscosities of
membrane and surrounding fluids, respectively.
Fournier {\it et al.}~\cite{four09} reported that higher friction between two
leaflets slows down membrane tubulation.
Thus, we expect a similar tendency in budding and tubulation.
Recently, Fujiwara and Yanagisawa~\cite{fujiwara2014generation,fujiwara2017liposomal}
reported that vesicles containing high
concentrations of macromolecules undergo bud or tube formation depending on the viscosity of the inner solution;
the membrane tube formation appears at the high viscosity, otherwise, bud
formation appears. The slow elastic relaxation due to the high viscosity
suppresses the bud formations, and enhances the tube formations.
We consider that the viscosity ratio of membrane and surrounding fluids
is also important in their experiments
as well as the viscosity ratio of the inner and outer fluids.

Although we focused on the effects of the viscosity ratio here,
the static membrane properties, bending rigidity and edge line tension, also modify the shape transformations.
Large bending rigidity and/or small edge tension suppress the transition from the BP to the spherical vesicle.
Small edge tension likely also enhances the BP formation owing to the reduction of the nucleation energy.
The bending rigidity can be reduced by EOs~\cite{nakagawa2015morphological}, and
the edge tension can be reduced by the addition of cone-shape surfactant molecules.

In living cells, lipid droplets are formed on the ER membrane~\cite{bell1981lipid}.
The hydrophilic segments of the lipids are removed by chemical reactions.
These reactions are similar to the hydrolysis reaction in our simulation.
The resulting hydrolyzed hydrophobic molecules assemble into clusters in the bilayer membrane.
We expect that a similar cluster formation plays a role in the initial lipid droplet formation.

In this work, the rupture of the vesicles is not observed due to the high edge line
tension of the bilayer. However, in the experiments conducted by Riske
{\it et al.}~\cite{riske2003diacylglycerol}, the rupture of liposomes occurs as
a result of the
injection of hydrolase enzyme. The coupling of the asymmetrical amphiphilic
molecular density and membrane rupture causes the inside-out inversion~\cite{nomura2001capabilities}.
Under low edge tension, competition between ruptures and BP formation may occur.

Here, we only consider amphiphilic molecules that form the bilayer.
Chemical reactions can change the shape of amphiphilic molecules from cylinder to cone or inverted-cone shapes.
In the experiments by Suzuki {\it et al.}~\cite{suzuki2009sparkling},
the molecular assembly changes their shapes from tubular micelles to vesicles.
In the future, it will be interesting to investigate the molecular mechanism of
these shape transformations involving non-bilayer structures.

\begin{acknowledgments}
  We would like to thank to K. Suzuki, T. Toyota, T. Sugawara, G. Gommper, and R. Hornung for helpful discussions.
  This work was supported by JSPS KAKENHI Grant Number JP25103010 and JP16J01728.
  The numerical calculations were partly carried out by SGI Altix ICE XA at ISSP
  Supercomputer Center, University of Tokyo.
\end{acknowledgments}

\begin{appendix} 
\section{Preparation of vesicles}\label{apdx:prep_ves}

We prepare the spherical vesicles in the following manner.
First, a spherical-cap-shaped bilayer membrane is centered in the simulation box,
and then the system is equilibrated during $500\tau \sim 2500\tau$.
A closed spherical membrane is thus obtained.
Some water particles inside the vesicle are then ejected to
tune the reduced volume $v$.
After that, the system is equilibrated for $5000\tau$.
For these system equilibrations, we use $\gamma_{ij}^* = 2$
to speed up the equilibration processes.
The system is then again equilibrated for $3000\tau$ using
the same setting $\gamma_{ij}$ as for the production runs.
After the above-mentioned equilibration processes,
we turn on the chemical reactions.

\section{Calculation of thickness of vesicles}\label{apdx:calc_thick}
We calculate the bilayer thickness $d$ of vesicles from the two layers of the
hydrophilic particles.
First, we extract the positions of hydrophilic particles in the bilayer membrane.
Next, we carry out a clustering analysis using depth-first search with search
radius $r_{s} = 0.25r_{\mathrm{cut}}$.
Two sets (clusters) of hydrophilic particles are obtained in most cases
before the large shape deformations:
One is the head-group of the inner leaflet, and
the other is the head-group of the outer leaflet. We define the local bilayer thickness as
\begin{align}
 d(i) = \min_{j \in C_{\mathrm{out}}} {r_{ij}} \, (\mathrm{for} \, i \in C_{\mathrm{in}}),
\end{align}
where $C_{\mathrm{in}}$ and $C_{\mathrm{out}}$ are the sets of hydrophilic
particles in the inner and outer leaflets, respectively.
The membrane thickness $d$ is defined as the mean value of $d(i)$.
The thickness inhomogeneity $\delta d$ defined as
\begin{align}
 \delta d = \sqrt{\dfrac{1}{n(C_{\mathrm{in}})} \sum_{i\in C_{\mathrm{in}}} (d(i) - d)^{2}},
\end{align}
where $n(C_{\mathrm{in}})$ is the number of particles included in the inner
leaflet.
\end{appendix}

% \bibliographystyle{apsrev}
% \bibliography{bpbd,add}

\end{document}